\documentclass[letterpaper, 10 pt, conference]{ieeeconf}  

\IEEEoverridecommandlockouts                              

\usepackage{cite}
\usepackage{amsmath,amssymb,amsfonts,bbold}
\usepackage{graphicx}
\usepackage{textcomp}
\usepackage{xcolor}
\usepackage{algpseudocode}
\usepackage{algorithm}

\usepackage{mathtools}
\mathtoolsset{showonlyrefs,showmanualtags}

\renewcommand{\baselinestretch}{0.97}

\newtheorem{remark}{Remark}[section]

\newcommand{\ve}{\boldsymbol}

\begin{document}


\title{A Metropolis-Adjusted Langevin Algorithm for Sampling Jeffreys Prior
\thanks{This work has been partially supported by the Swedish Research Council under contract number 2023-05170, and by the Wallenberg AI, Autonomous Systems and Software Program (WASP) funded by the Knut and Alice Wallenberg Foundation. The authors are with the Division of Decision and Control Systems, KTH Royal Institute of Technology, SE-100 44 Stockholm, Sweden. Emails: \texttt{\{yibos, blak, crro\}@kth.se}}}
\author{Yibo Shi, Braghadeesh Lakshminarayanan and Cristian R. Rojas}


\maketitle

\begin{abstract}
Inference and estimation are fundamental in statistics, system identification, and machine learning. When prior knowledge about the system is available, Bayesian analysis provides a natural framework for encoding it through a prior distribution. In practice, such knowledge is often too vague to specify a full prior distribution, motivating the use of default “uninformative’’ priors that minimize subjective bias.
Jeffreys prior is an appealing uninformative prior because: (i) it is invariant under any re-parameterization of the model, (ii) it encodes the intrinsic geometric structure of the parameter space through the Fisher information matrix, which in turn enhances the diversity of parameter samples. Despite these benefits, drawing samples from Jeffreys prior is challenging.
In this paper, we develop a general sampling scheme using the Metropolis-Adjusted Langevin Algorithm that enables sampling of parameter values from Jeffreys prior; the method extends naturally to nonlinear state–space models. The resulting samples can be directly used in sampling-based system identification methods and Bayesian experiment design, providing an objective, information-geometric description of parameter uncertainty. Several numerical examples demonstrate the efficiency and accuracy of the proposed scheme.
\end{abstract}


\section{Introduction}

Mathematical models are essential tools for analyzing, predicting, and controlling complex physical processes. System identification is a discipline that deals with the construction of such models from experimental input-output data~\cite{Ljung:99,Soderstrom-Stoica-89}. A central task within system identification is parameter estimation: given a chosen (or presumed) model structure, the goal is to infer its unknown parameters from observed data.
Classical approaches include Prediction Error Methods~\cite{Ljung:99}, Instrumental Variables~\cite{Soderstrom-Stoica-83}, Subspace Methods~\cite{van2012subspace}, and Maximum Likelihood Estimation~\cite{caines1975note}. These methods typically yield point estimates that are often consistent and even asymptotically efficient. However, as the dataset used for estimation is inherently finite, there remains a level of uncertainty in the resulting parameter estimates that must be addressed. To rigorously capture this uncertainty, Bayesian parameter estimation provides a systematic probabilistic framework, enhancing inference reliability and robustness.

In the Bayesian framework~\cite{Bernardo-Smith-09}, parameter estimation is updated via Bayes’ theorem when observational data become available, producing a posterior distribution that reflects the newly inferred probabilities of the model parameters.
Mathematically, the posterior's log-density can be decomposed into: (i) the log-likelihood, which measures how well parameters explain observations, and (ii) the log-prior, which encodes initial beliefs. This formulation reveals that the prior distribution inherently acts as a regularization term~\cite{pillonetto2022regularized,figueiredo2003adaptive}. For instance, Laplace priors enforce sparsity via $l_1$ regularization, while Gaussian priors impose $l_2$-type constraints on parameter magnitude. Consequently, the choice of prior directly influences the balance between fitting the data and imposing constraints during posterior inference, shaping the final parameter estimates.

Among all possible prior probability densities, Jeffreys prior~\cite{jeffreys1946invariant,Jeffreys-61} offers several advantages: (i) it has minimal subjective influence, making it a suitable choice in the absence of strong domain-specific information; (ii) its form remains consistent under invertible transformations of the parameter, thus it ensures that posterior updates remain invariant to arbitrary parameter transformations, avoiding the bias introduced by coordinate choices; and (iii) the prior is proportional to the square root of the determinant of the Fisher Information Matrix (FIM), which encodes the intrinsic geometric structure of the parameter space. This last property is particularly valuable in data-driven parameter estimation~\cite{garatti2013new}, where the induced Riemannian geometry facilitates sampling diverse parameter values. Such diversity enhances the robustness of machine learning models trained on these samples, as the parameters reflect distinct regions. Despite these advantages, Jeffreys prior faces practical limitations. Computing the FIM in closed form may be intractable or prohibitively expensive for complex or high-dimensional models. While numerical approximations can yield partial FIM estimates, they introduce additional uncertainty in the sampling process: the derived potential function is subject to estimation errors, potentially degrading performance in naive sampling schemes. Furthermore, Jeffreys prior’s normalizing constant is rarely known explicitly, even if the parameter space is bounded and the prior is integrable, complicating direct sampling.




These difficulties are typically addressed through Markov chain Monte Carlo (MCMC) methods, which are widely used for sampling from challenging or high-dimensional distributions \cite{robert1999monte}. Among these, Langevin-based Monte Carlo (LMC)~\cite{roberts1996exponential} employs gradient information to guide proposals efficiently in parameter space.
Several variants have been proposed to address some specific challenges of LMC. For instance, {Constrained Ensemble LMC}~\cite{ding2022constrained} ensures physically valid sample updates by projecting the samples onto constraint sets at each iteration; the {Metropolis--Adjusted Langevin Algorithm} (MALA) augments LMC with a Metropolis--Hastings accept/reject step, correcting discretization errors and guaranteeing exact convergence to the target distribution regardless of moderate step-size mis-specifications~\cite{roberts1996exponential,roberts2002langevin}. Among the LMC variants, MALA offers a simple way to handle constraints and correct for discretization errors while preserving the geometric advantages of Jeffreys prior via a gradient-based proposal mechanism. 

In this paper, we introduce a MALA-based scheme for sampling Jeffreys prior in two distinct scenarios: (i) when the FIM can be derived analytically, and (ii) when an analytical form of the FIM is unavailable, but the score function can be estimated via particle filtering (PF)~\cite{dahlin2014particle,valenzuela2017robust}, allowing us to compute an approximate FIM that is then used within MALA. Furthermore, we demonstrate an application of Jeffreys prior to generate a diverse set of parameter samples to enhance the performance of a data-driven estimator, by providing improved estimates of the parameters of a model. Our main contributions include:
\begin{itemize}
    \item A MALA-based sampling approach tailored specifically to Jeffreys prior;
    \item Extension of this scheme to nonlinear dynamical systems through PF-based FIM estimation;
    \item Application of sampled Jeffreys prior to promote informative and diverse parameter sets for data-driven estimation methods, thus improving their performance.
\end{itemize}

The remainder of the paper is organized as follows: in Section~\ref{sec: prob_stat}, we define the problem statement. Section~\ref{sec: proposed method} introduces the proposed algorithm to sample from Jeffreys prior, while Section~\ref{sec: numerical study} provides several numerical illustrations of our method. Finally, we conclude the paper in Section~\ref{sec: conclusion}.

\section{Problem statement} \label{sec: prob_stat}

Consider a family of probability distributions $\{p( \cdot ; \ve{\theta})\colon \ve{\theta} \in \Theta\}$ defined on a sample space $\mathcal{Y}$, where $\Theta \subseteq \mathbb{R}^d$ is the parameter space.\footnote{Throughout this paper, we use boldface fonts (e.g., $\ve\theta$) to refer to vector or matrix variables and normal fonts (e.g., $\theta$) for scalar variables.} The FIM at a given parameter value $\ve \theta$, denoted by $\ve{J}_{\ve \theta}$, is defined as 
\begin{equation}
\ve{J}_{\ve \theta}
\;=\;
\mathbb{E}_{y \sim p(\cdot; \ve{\theta})}
\!\Bigl[
\nabla_{\ve \theta} \ln p(y; \ve{\theta})\,
\nabla_{\ve \theta^{\top}} \ln p(y; \ve{\theta})
\Bigr],
\label{eq:fisher-info-def}
\end{equation}
where $y\in \mathcal{Y}$ denotes the observations, $\nabla_{\ve \theta}$ denotes the gradient with respect to $\ve \theta$, and $\mathbb{E}_{y \sim p(\cdot;\,\boldsymbol{\theta})}[\cdot] =\int_{\mathcal{Y}} \bigl[\cdot\bigr]\, p(y;\,\boldsymbol{\theta})\,dy.$

Intuitively, $\ve{J}_{\ve \theta}$ measures the local sensitivity of the log-likelihood to changes in $\ve{\theta}$ and can be viewed as a Riemannian metric on the parameter space $\Theta$ \cite{boothby2003introduction}.

In practice, computing \(\boldsymbol{J}_{\boldsymbol{\theta}}\) for many complex or high-dimensional models can be analytically intractable or computationally prohibitive, especially when the distributions lack closed-form expressions (see, e.g., \cite{valenzuela2017robust} for an example in nonlinear dynamical systems). Consequently, one often resorts to {approximate} or {Monte Carlo} methods to estimate \(\boldsymbol{J}_{\boldsymbol{\theta}}\). For instance, given a finite dataset \(\{y_i\}_{i=1}^n\subset\mathcal{Y}\), one can compute sample-based estimates  $\ve{\hat{J}}_{\ve \theta}$ by replacing the expectation in \eqref{eq:fisher-info-def} with an empirical average \cite{valenzuela2017robust}.


{Jeffreys prior}~\cite{jeffreys1946invariant} is an uninformative prior distribution on the parameter space $\Theta$. 
Concretely, Jeffreys prior $\pi(\ve{\theta})$ is defined (up to a constant factor) by
\begin{equation}
\pi(\ve{\theta})
\;\propto\;
\sqrt{\det(\ve{J}_{\ve \theta})},
\label{eq:jeffreys-prior}
\end{equation}
where 
$\det(\cdot)$ denotes the determinant function.

One key advantage of Jeffreys prior over alternative uninformative priors is its {reparameterization invariance}. If one changes variables from $\ve \theta$ to $\ve \phi = g(\ve \theta)$ via a smooth invertible transformation, the associated Jeffreys prior adjusts automatically, preserving the same degree of ``non-informativeness'' in the new parameter space. This property makes Jeffreys prior a canonical choice when one wishes to impose as little subjective structure as possible.

Most importantly, Jeffreys prior's explicit dependence on the FIM \(\ve J_{\ve \theta}\) has a Riemannian geometric interpretation. Indeed, as mentioned above, a suitable Riemannian metric on $\Theta$ is given by the FIM \(\ve J_{\ve \theta}\). In particular, \(\Delta \ve \theta^T \ve J(\ve \theta) \Delta \ve \theta\) is a good measure of how ``different" the probability distributions $p(\cdot;\ve \theta)$ and $p(\cdot; \ve \theta + \Delta \ve \theta)$ are, for $\Delta \ve \theta$ sufficiently small~\cite{Amari-16}. Since such a metric induces a natural volume element
of the form $d\mathrm{vol} = \sqrt{\det(\ve{J}_{\ve{\theta}})} \, d\theta^1 \wedge \cdots \wedge d\theta^d$~\cite[pp.~233]{boothby2003introduction}, it helps us to distribute samples from $\Theta$ according to this volume element so that these samples are well distributed in the parameter space.  

Sampling from Jeffreys prior is thus highly relevant in system identification, machine learning, and statistics, but poses two main challenges: (i) for complex models, computing or approximating the FIM can introduce uncertainty, and (ii) the normalizing constant of Jeffreys prior is rarely known, complicating direct sampling.
The objective of this paper is to address these challenges by proposing an LMC approach that generates Markovian samples whose stationary distribution corresponds precisely to Jeffreys prior $\pi(\ve \theta)$. 


\section{Proposed Method} \label{sec: proposed method}


In this section, we first describe two standard sampling methods: LMC, and its improved variant, MALA.
Then, in Subsection~\ref{sec:result}, we introduce a novel MALA-based method for sampling from Jeffreys prior, which accommodates two scenarios: one where the FIM can be computed analytically, and the other where it must be approximated. 

\subsection{Langevin-based Monte Carlo}

The LMC provides an efficient, gradient-based framework for exploring the Jeffreys prior distribution. In this setting, we generate samples distributed according to 
\begin{equation}
\ve\theta \sim \pi(\ve\theta) \propto \exp(-V(\ve\theta))
\label{eq:jeffreys-samples}
\end{equation}
by simulating the following Langevin stochastic differential equation (SDE)~\cite{Karatzas-Shreve-91}:
\begin{equation}
d\ve\theta_{t}= -\nabla_{\ve \theta} V(\ve\theta_t)dt + \sqrt{2}d\ve{w}_{t}, 
\label{eq:langevin-sde}
\end{equation}
where $\ve{w}_{t}$ denotes standard Brownian motion in $\mathbb{R}^d$, and $V\colon \mathbb{R}^d \rightarrow \mathbb{R}$ is a differentiable potential function. 

A standard numerical method to simulate \eqref{eq:langevin-sde} numerically is the {Euler--Maruyama} method~\cite{Higham-Kloeden-21}. Discretizing time in steps of size \(\tau\), we obtain
\begin{align}
\ve\theta_{i+1} &= \ve\theta_i - \tau\nabla_{\ve \theta} V(\ve\theta_i)+\sqrt{2\tau}\bigl(\ve{w}_{i+1} - \ve{w}_{i}\bigr), \\
             &= \ve\theta_i - \tau\nabla_{\ve \theta} V(\ve\theta_i)+\sqrt{2\tau}\ve\xi_i,
\label{eq:ULA}
\end{align}
where \(\ve\xi_i \sim \mathcal{N}(\ve{0}_d,\ve{I}_d)\). The discretization step size $\tau$ can be fixed or adapted during simulation.

To verify that the SDE in \eqref{eq:langevin-sde} samples from the Jeffreys prior defined in \eqref{eq:jeffreys-samples}, we use the corresponding {Fokker--Planck} equation~\cite{Karatzas-Shreve-91}. For simplicity of illustration, we consider the one-dimensional case with parameter \(\theta\). The Fokker--Planck equation~\cite{Karatzas-Shreve-91} associated with the Langevin SDE
\begin{equation*}
d\theta_t = -\frac{dV(\theta_t)}{d\theta}\,dt + \sqrt{2}\,dw_t
\label{eq:langevin-sde-1d}
\end{equation*}
describes the evolution of the probability density \(p(\theta,t)\) as
\begin{equation*}
\frac{\partial p(\theta,t)}{\partial t}
\;=\;
-\frac{\partial}{\partial \theta}\Bigl[p(\theta,t)\,\frac{dV(\theta)}{d\theta}\Bigr]
\;+\;
\frac{\partial^2 p(\theta,t)}{\partial \theta^2}.
\label{eq:fokker-planck}
\end{equation*}

In the steady-state regime \(\partial p(\theta,t)/\partial t=0\), the probability density reaches a stationary distribution denoted by \(p_{\infty}(\theta)\). Imposing zero probability flux at the boundaries (or as \(|\theta|\to\infty\)), we obtain $p_{\infty}(\theta) \;\propto\; \exp\bigl(-V(\theta)\bigr)$. Hence, the normalized stationary distribution is
\begin{equation} \label{eq:steady-state}
p_{\infty}(\theta)
\;=\;
\frac{\exp\bigl(-V(\theta)\bigr)}{Z},
\end{equation}
where \(Z\in\mathbb{R^+}\) is a normalizing constant.

If we specifically define the potential function as
\begin{align}\label{eq:V_pi}
    V(\theta) = -\frac{1}{2}\ln ( J_\theta),
\end{align}
then it follows from \eqref{eq:steady-state} that
$
p_{\infty}(\theta)
\;
\propto\;
\pi(\theta).$
Thus, the stationary distribution associated with the Langevin SDE \eqref{eq:langevin-sde} coincides exactly with Jeffreys prior. 

The update rule in \eqref{eq:ULA} to sample from $\pi(\ve\theta)$ leads to the {Unadjusted Langevin Algorithm} (ULA), which iterates \eqref{eq:ULA} to generate the samples $\ve\theta_i$ for $i = 0,1,2,\dots$, whose distribution approximates $\pi(\ve\theta)$ after a sufficient burn-in period. However, ULA is highly sensitive to the choice of step size \(\tau\). If \(\tau\) is excessively large, discretization errors introduce bias into the invariant distribution or even cause divergence; if too small, the chain suffers slow mixing, resulting in computational inefficiency.
Moreover, one often requires the sampled parameter $\ve\theta_i$ to stay in a constrained parameter space $\Theta_c \subset \Theta$ due to physical limitations or prior knowledge. ULA lacks a mechanism to enforce these constraints. 
In the following subsection, we describe the MALA framework, which addresses these limitations. 

\subsection{Metropolis-Adjusted Langevin Algorithm}

MALA~\cite{roberts2002langevin} improves upon ULA by introducing a Metropolis–Hastings accept/reject step that compensates for the errors induced by the discretization. Note that in \eqref{eq:ULA}, the proposed variable $\ve\theta_{i+1}$ follows the Gaussian distribution
\begin{align}
    \ve\theta_{i+1} \sim \mathcal{N}(\ve\theta_i - \tau\nabla_{\ve \theta} V(\ve\theta_i), 2\tau).
\end{align}
The proposal density can thus be explicitly written as
\begin{align}
q\bigl(\ve\theta_{i+1}\mid \ve\theta_i\bigr)
\propto
\exp\Bigl(
-\frac{\|\,\ve\theta_{i+1} - \ve\theta_i +  \tau\nabla_{\ve \theta} V(\ve\theta_i)\|_2^2}{4\,\tau}
\Bigr).
\end{align}
MALA accepts a given proposed sample $\ve\theta'$ with probability
\begin{align*}
\rho^{\text{MALA}}(\ve\theta', \ve\theta_i)
\;=\;
\min \Bigl\{
1,\;
\frac{\exp(-V(\ve\theta'))\,q\bigl(\ve\theta_i\mid \ve\theta'\bigr)}
     {\exp(-V(\ve\theta_i))\,q\bigl(\ve\theta'\mid \ve\theta_i\bigr)}
\Bigr\}.
\end{align*}
If the proposal is accepted, the chain advances as $\ve\theta_{i+1} = \ve\theta'$; otherwise, it remains at the current position, $\ve\theta_{i+1} = \ve\theta_i$. This ensures that the chain remains exactly invariant with respect to $\pi(\ve\theta)$ under mild conditions, yielding more robust sampling compared to ULA \cite{roberts2002langevin}. Furthermore, MALA inherently accommodates constrained parameter spaces $\ve{\theta} \in \Theta_{\text{c}}$ by modifying the acceptance probability as follows:
$$
\rho_c^{\text{MALA}}\left(\boldsymbol{\theta}^{\prime}, \boldsymbol{\theta}_i\right)= \begin{cases}\rho^{\text{MALA}}\left(\boldsymbol{\theta}^{\prime}, \boldsymbol{\theta}_i\right), & \text{if } \boldsymbol{\theta}' \in \Theta_c, \\ 0, & \text{otherwise } .\end{cases}
$$


\subsection{Sampling with Estimated FIM}\label{sec:result}


If an analytical form of the FIM is available, we can directly define the multivariate potential function
\begin{align}\label{eq:V_pi_vec}
    V(\ve\theta) = -\frac{1}{2}\ln \det (\ve J_{\ve\theta}),
\end{align}
which is a generalization of~\eqref{eq:V_pi}, with gradient
\begin{align}\label{eq:V_grad}
    \nabla_{\ve \theta}V(\ve\theta) = - \frac{1}{2}\text{tr}\left[\ve J_{\ve\theta}^{-1} \frac{\partial \ve J_{\ve\theta}}{\partial \ve\theta} \right],
\end{align}
to sample from Jeffreys prior via MALA. 

For many problems, such as nonlinear state-space (NLSS) models, closed-form expressions for FIMs are not available. In such cases, one must resort to numerical approximations. In this paper, we adopt a PF–based approach to estimate the FIM, following \cite{valenzuela2017robust}, whereby {Forward Filtering--Backward Smoothing} (FFBSm) provides an unbiased Monte Carlo approximation of the score function, and consequently, the FIM. Despite the inherent stochasticity of particle filters, the resulting estimates are consistent and converge to the true FIM under standard regularity conditions as the number of particles increases.

\begin{remark}
For a concrete illustration of PF–based FIM estimation within our framework, we refer the reader to the NLSS model simulation in Section~\ref{sec: numerical study}.
\end{remark}



Given the estimated FIM $\ve{\hat{J}}_{\ve\theta}$, the gradient ${\partial \ve J_{\ve\theta}}/{\partial \ve\theta}$ can be approximated by the {one-point unbiased estimate}. Specifically, we introduce a random perturbation \(\ve\mu \sim \mathcal{N}(\ve{0}_d,\ve{I}_d)\) and approximate the derivative as
\begin{align}\label{eq:onepoint}
    \frac{\partial \ve J_{\ve\theta}}{\partial \theta_j} \approx \frac{\mu_j}{\delta}\Bigl(\ve{\hat{J}}_{\ve\theta+\delta\ve\mu}- \ve{\hat{J}}_{\ve\theta}\Bigr), \quad j = 1, \dots, d,
\end{align}
where \(\delta>0\) is a small step size, and $\ve{\hat{J}}_{\ve\theta+\delta\ve\mu}$ is computed using the same estimation procedure as for $\ve{\hat{J}}_{\ve\theta}$. This estimator is unbiased (for $\delta \to 0$) and substantially faster compared to coordinate-wise finite differences. 



\begin{remark}
Although the use of the approximate gradient in~\eqref{eq:onepoint} might introduce errors, the Metropolis--Hastings step compensates for these inaccuracies and ensures that the target distribution \(\pi(\ve\theta)\) is still preserved \cite{andrieu2009pseudo}. Standard results in MCMC theory~\cite{roberts1996exponential, roberts2002langevin} guarantee that, as long as the gradient estimator is unbiased and its noise is properly accounted for by the acceptance step, the resulting Markov chain will have \(\pi(\ve\theta)\) as its stationary distribution. 
\end{remark}

Algorithm \ref{alg:MALA} integrates the PF–based FIM estimation with the one-point gradient approximation within MALA. In this way, even when the FIM is only approximately available, our approach guarantees convergence to Jeffreys prior. 



\begin{algorithm}[h]
\caption{Sample from Jeffreys Prior Distribution}
\label{alg:MALA}
\begin{algorithmic}[1]
\Require Initial guess $\ve\theta_0$, constraint set $\Theta_{\text{c}}$, step size $\tau$, number of iterations $N$, 
  small step size $\delta > 0$
\For {$n = 0, 1, \dots, N-1$}
  \State Compute or estimate $\ve J_{\ve \theta_n}$
  \State Compute $\nabla_{\ve\theta}V(\ve\theta_n)$ using~\eqref{eq:V_grad} \textbf{or} run 
  \State \hspace{5mm} Draw a random direction $\ve \mu \sim \mathcal{N}(\ve{0}_d, \ve{I}_d)$ 
  \State \hspace{5mm} Estimate $\nabla_{\ve\theta}V(\ve\theta)$ using~\eqref{eq:onepoint}
  \State \hspace{5mm} Estimate $\ve J_{\ve\theta+\delta\ve\mu}$ 
  \State \hspace{5mm} Compute $\nabla_{\ve\theta}V(\ve\theta_n)$ using~\eqref{eq:V_grad}
  \State Sample \(\ve\xi_n \sim \mathcal{N}(\ve{0}_d,\ve{I}_d)\) and $U \sim \mathcal{U}(0,1)$
  \State \textbf{Propose} 
    \[
      \ve \theta' 
      \leftarrow \ve\theta_n - \tau\nabla_{\theta} V(\ve\theta_n)+\sqrt{2\tau}\ve\xi_n,
    \]
    \[
    \rho_n \leftarrow
    \begin{cases}
        \rho^{\text{MALA}}(\ve\theta', \ve\theta_n) , & \text{If } \boldsymbol{\theta}' \in \Theta_c,\\
        0, & \text{Otherwise}.
      \end{cases}
    \]
  \State \textbf{Accept/reject step}
    \[
      \ve {\theta}_{n+1}
      \;\leftarrow\;
      \begin{cases}
        \ve \theta' , & \text{If } U<\rho_n,\\
        \ve \theta_n, & \text{Otherwise}.
      \end{cases}
    \]
\EndFor
\State \Return $\{\ve\theta_n\}_{n=1}^{N}$
\end{algorithmic}
\end{algorithm}

\section{Numerical Illustrations} \label{sec: numerical study}

In this section, we present three numerical examples:

\begin{enumerate} \itemsep 0pt
    \item Sanity check: a simple example in which Jeffreys prior can be computed exactly, verifying that Algorithm \ref{alg:MALA} converges to the correct distribution;
    \item NLSS system: a sampling example where the FIM is estimated via the PF method, showing that Algorithm~\ref{alg:MALA} remains effective with estimated $\ve J_{\ve\theta}$;
    \item Parameter estimation: a practical setting in which Jeffreys prior sampling outperforms a uniform prior in data-driven parameter estimation performance. 
\end{enumerate}


\subsection{Verification of the Sampling Procedure}

\begin{figure*}[h]
    \centering
    \includegraphics[width=1\linewidth]{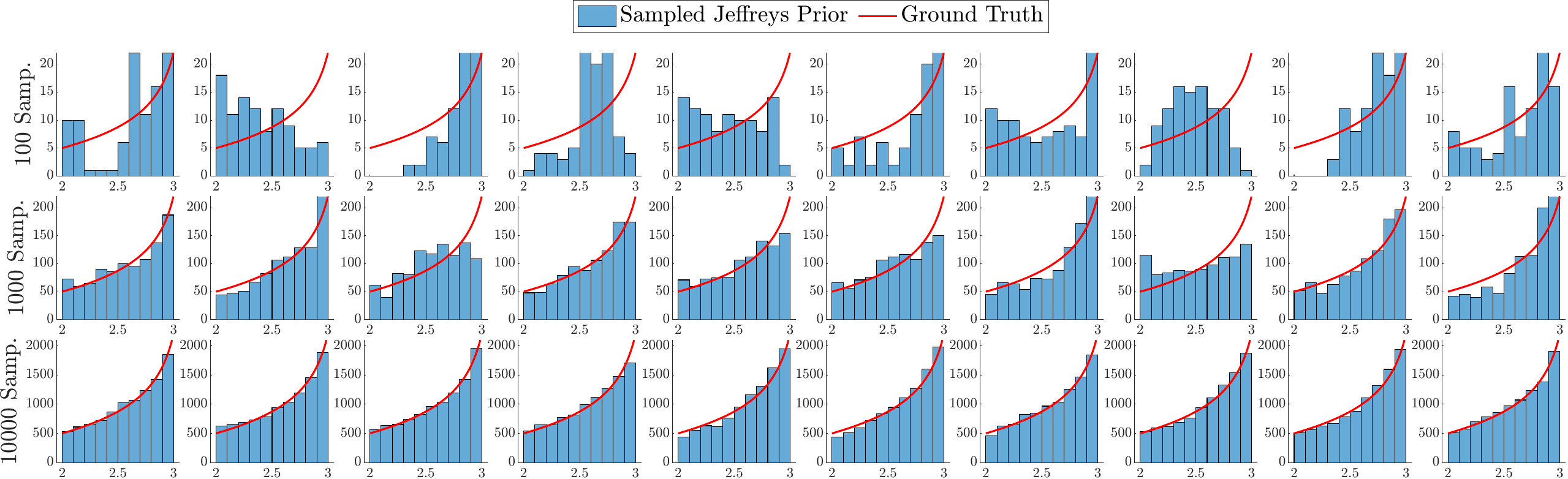}
    \caption{Histograms of 100 (top), 1000 (middle), and 10000 (bottom) samples of Jeffreys prior and the shape of the true Jeffreys prior (red curve) from the Coin-Bending Model. Each set of sampling experiments contains ten different realizations.}
    \label{fig:coin}
\end{figure*}

\subsubsection{Experiment setup}

In this experiment, we apply Algorithm~\ref{alg:MALA} to sample from Jeffreys prior for a Coin-Bending Model~\cite{ly2017tutorial}, where the probability $q(\varphi)$ of obtaining heads depends on the bending angle $\varphi$ according to 
$$
q(\varphi)=\frac{1}{2}+\frac{1}{2}\left(\frac{\varphi}{\pi}\right)^3.
$$
Each coin toss yields a binary random variable $Y$, where
$Y = 1$ with probability $\frac{1}{2}+\frac{1}{2}\left(\frac{\varphi}{\pi}\right)^3$, and $Y = 0$ otherwise,
%
with 1 denoting heads. Its score function is
$$
\nabla_\varphi \ln p(Y ; \varphi)=\mathbb{1}\{Y=1\} p_1-\mathbb{1}\{Y=0\} p_0,
$$
with
$$ 
p_1 = \frac{\frac{3}{\pi}\left(\frac{\varphi}{\pi}\right)^2}{1+\left(\frac{\varphi}{\pi}\right)^3}, \quad p_0 = \frac{\frac{3}{\pi}\left(\frac{\varphi}{\pi}\right)^2}{1-\left(\frac{\varphi}{\pi}\right)^3}.
$$
The parameter to be estimated is $\varphi$. Given $n$ independent coin tosses resulting in observations $\boldsymbol{y} \in\{0,1\}^n$, the information $J_{\varphi}$ per sample can be approximated, if $n$ is sufficiently large, by
\begin{align*}
    \hat{J}_{\varphi}&=\frac{\mathbf{1}^T \boldsymbol{y}}{n}p_1^2+\frac{n-\mathbf{1}^T \boldsymbol{y}}{n}p_0^2.
\end{align*}
Assuming consistency as $n \rightarrow \infty$, we treat $\pi(\varphi) \propto \sqrt{\hat{J}_{\varphi}}$ as the ground truth, and define the potential $V(\varphi)=-\frac{1}{2} \ln \hat{J}_\varphi$ whose gradient given explicitly by
\begin{align*}\label{eq:coindV}
    \nabla V(\varphi)=-\frac{1}{2} \hat{J}_{\varphi}^{-1} \frac{d\hat{J}_{\varphi}}{d \varphi}.
\end{align*}

We run Algorithm~\ref{alg:MALA} to generate samples from $\pi(\varphi)$ within the constrained set $\varphi \in \Phi_c = [2, 3]$. Three experiments are performed with $N \in \{100,\ 1000,\ 10000\}$ samples, each repeated for 10 independent realizations. The resulting empirical distributions are then compared with the estimated Jeffreys prior to verify convergence and consistency. 

\subsubsection{Simulation results}
The simulation results are shown in Fig. \ref{fig:coin}, which shows that the empirical distribution of samples obtained by our algorithm matches the theoretical Jeffreys prior as the number of samples increases (from Row 1 to Row 3). Moreover, only a relatively small sample size (e.g., 1,000 samples) yields a very good match to the exact Jeffreys prior, indicating rapid convergence. Additionally, the difference in the sampled realizations (for $N = 100$) reveals the stochastic nature of the chain, demonstrating good mixing behavior and adequate exploration of the parameter space.

This example provides a clear validation of our proposed MALA-based sampler, as it allows direct comparison between the empirical and exact prior distributions. Furthermore, this one-dimensional scenario enables straightforward assessment of mixing quality and acceptance rates. 

\subsection{Sampling Jeffreys Prior for a Dynamical System}

\subsubsection{Experiment setup}
In this example, we apply Algorithm~\ref{alg:MALA} on an NLSS model, where the FIM is numerically approximated using a PF. Specifically, we consider the Hull--White stochastic volatility (SV) model \cite{hull1987pricing,dahlin2014particle}
\begin{align*}
    x_{t+1} \mid x_t \sim \mathcal{N}\left(\varphi x_t + \rho u_t, \sigma_v^2\right),\quad
    y_t \mid x_t \sim \mathcal{N}\left(0,\beta^2 e^{x_t}\right),
\end{align*}
where the control input $u_t$ is white noise of standard normal distribution. Our goal is to sample the Jeffreys prior $\pi(\varphi)$ for the system parameter $\varphi$ corresponding to the system dynamics, constrained to the feasible region $\Phi_c = [0.3, 0.9]$, ensuring system stability. For simulation, the remaining model parameters are set as $[\rho,\ \sigma_v,\ \beta] = [0.2,\ 0.5,\ 0.7]$, and the system is simulated for $T=1000$ time steps. 

Since the FIM $J(\varphi)$ lacks an analytical form for this model, we estimate it using a PF–based method from \cite{valenzuela2017robust}, as described in Section~\ref{sec: proposed method}, employing $N_p=1000$ particles and averaging multiple Monte Carlo runs to obtain a reliable estimate $\hat{J}(\varphi)$. Accordingly, we compute the estimated shape of Jeffreys prior $\pi(\varphi)$ and compare it with the distribution of the generated samples.

We define the potential function as $V(\varphi) = -\frac{1}{2}\ln \hat{J}(\varphi)$ and approximate its gradient $\widehat{\nabla V}(\varphi)$ using~\eqref{eq:V_grad} and~\eqref{eq:onepoint}.

Finally, Algorithm~\ref{alg:MALA} is run for $N = 10000$ iterations with step size \(\tau = 0.05\), and the generated samples are compared against the estimated shape of Jeffreys prior. 

\subsubsection{Simulation results}

\begin{figure}[t]
    \centering
    \includegraphics[width=1\linewidth]{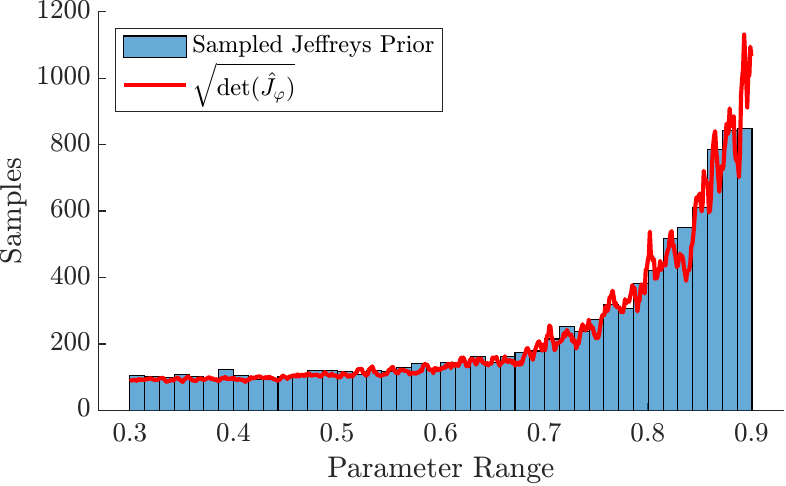}
    \caption{Histogram of 10000 samples of Jeffreys prior and the shape of the PF-estimated Jeffreys prior (red curve) from the SV Model.}
    \label{fig:SV}
\end{figure}

A histogram of the $10000$ generated samples is shown in Fig.~\ref{fig:SV}, overlaid with the estimated shape of Jeffreys prior $\pi(\varphi)$ computed from the PF estimation. 

As expected, the estimated Jeffreys prior assigns higher probability near the boundary \(\varphi = 0.9\), indicating greater parameter sensitivity, and lower probability near $\varphi = 0.3$.



Around $\varphi = 0.3$, the influence of \(x_t\) on the evolution of the state is relatively small; the dynamics are then dominated by the input \(u_t\) and the process noise \(v_t\), making the likelihood less sensitive to changes in $\varphi$. 
In contrast, around $\varphi = 0.9$, a small perturbation in \(\varphi\) produces significant changes in the predicted states and, hence, in the likelihood function, leading to a higher FIM. Thus, 
the PF-based FIM estimation is consistent with theoretical expectations.

Moreover, the close agreement observed in Fig.~\ref{fig:SV} between the histogram of generated samples and the estimated Jeffreys prior verifies that our proposed algorithm reliably samples from Jeffreys prior even when the FIM is numerically approximated. Overall, the results from experiments 1 and 2 demonstrate the validity of our sampling framework.



\subsection{An Application of Jeffreys Prior to Parameter Estimation}

\begin{figure*}[h]
    \centering
    \includegraphics[width=1\linewidth]{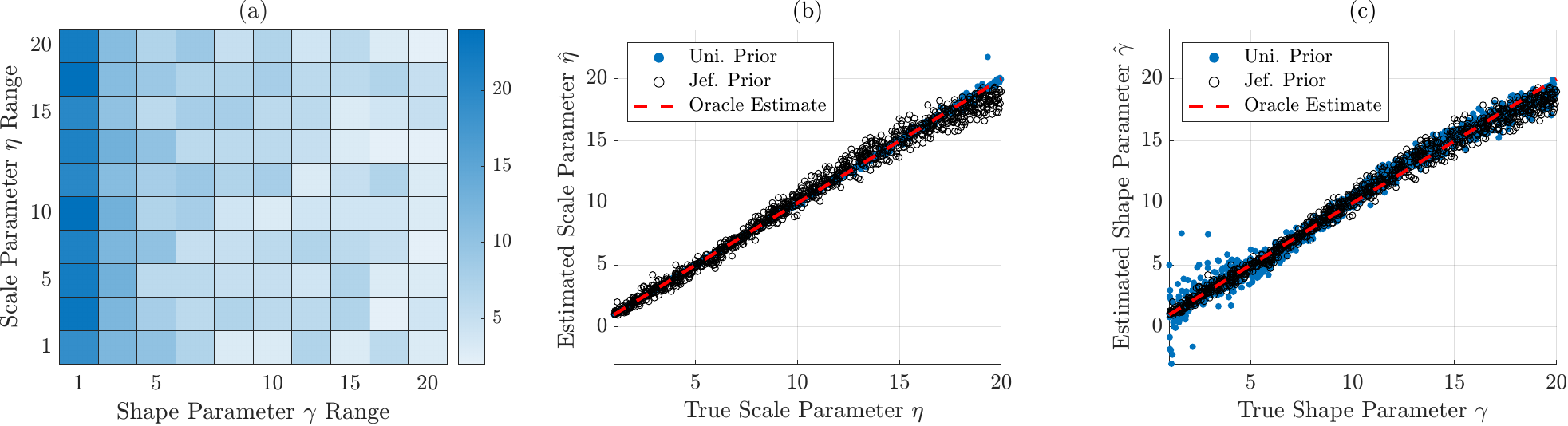}
    \caption{(a) Heatmap of 1000 samples from Jeffreys prior distribution of the scale ($\eta$) and shape ($\gamma$) parameters from a Weibull distribution; (b) Scatter plot of estimated scale parameter $\hat{\eta}$ based on the uniform prior and Jeffreys prior vs. its true value; (c) Scatter plot of estimated shape parameter $\hat{\gamma}$ based on the uniform prior and Jeffreys prior vs. its true value. The red dashed line corresponds to an oracle estimate, which knows the true value of the parameter.}
    \label{fig:weibull}
\end{figure*}

\subsubsection{Experiment setup}

In this example, we illustrate the advantages of sampling from Jeffreys prior within the Two-Stage (TS) estimation framework, using parameter estimation for a Weibull distribution as a test case. TS provides an effective validation scenario, although the framework itself is well-established. In TS, given a parametric model $p(\cdot;\boldsymbol{\theta})$ with parameters $\ve \theta \in \Theta \subseteq \mathbb{R}^d$, one constructs a synthetic training dataset of parameter-data pairs $\{ (\boldsymbol{\theta}_i, \boldsymbol{y}_i) \}_{i=1}^{M_\theta}$, where data samples $\boldsymbol{y}_i$ are generated from $p(\cdot;\boldsymbol{\theta}_i)$. A supervised learning method then builds an estimator of the form $\hat{\boldsymbol{\theta}}(\boldsymbol{y}) = \boldsymbol{g}(\boldsymbol{h}(\boldsymbol{y}))$, with a fixed compression function $\boldsymbol{h}$ and a regression-based function $\boldsymbol{g}$. For additional details,
see \cite{lakshminarayanan2022statistical}.

In the context of TS, the distribution of the synthetic parameter samples is crucial. We advocate sampling $\boldsymbol{\theta}$ according to Jeffreys prior, since $\pi(\boldsymbol{\theta})$ naturally reflects the local identifiability structure of the model, thus producing synthetic data which is most informative about the parameters. In contrast, a uniform prior may not account for how informative different regions of the parameter space are, potentially leading to suboptimal training datasets.
%
%

As a numerical example, we consider the Weibull distribution, widely used in reliability engineering, whose probability density function, parameterized by $\boldsymbol{\theta} = [\eta, \gamma]^T$, is given by
\[
f(A;\eta,\gamma) = \frac{\gamma}{\eta} \left(\frac{A}{\eta}\right)^{\gamma-1} \exp\left[-\left(\frac{A}{\eta}\right)^\gamma\right], \quad A \ge 0,
\]
where $\eta > 0$ and $\gamma > 0$ are the scale and shape parameters. 

\subsubsection{Simulation results}  
We validate the TS estimators trained under uniform and Jeffreys priors using a validation set consisting of $1000$ parameter points $\boldsymbol{\theta}_\ell = [\eta_\ell, \gamma_\ell]^T$ uniformly sampled from $[1,20]\times[1,20]$. For each $\boldsymbol{\theta}_\ell$ synthetic data $\{y^i_\ell\}_{i=1}^M$ generated from the Weibull model. We evaluate two classes of TS estimators based on the samples from the uniform and Jeffreys priors respectively.

Fig.~\ref{fig:weibull}(a) shows the sampled Jeffreys distribution in the $(\gamma,\eta)$ parameter space. Notably, Jeffreys prior emphasizes lower values of $\gamma$ while remaining relatively uniform along $\eta$, aligning well with the Weibull model’s FIM structure, which indicates higher sensitivity at smaller $\gamma$.

Estimation performances for $\eta$ and $\gamma$ are compared in Figs.~\ref{fig:weibull}(b) and \ref{fig:weibull}(c), respectively. In Fig.~\ref{fig:weibull}(b), both uniform and Jeffreys-based estimators produce accurate and similar results of the scale parameter across its entire range, because Jeffreys prior is nearly uniform in $\eta$ and thus provides coverage comparable to the uniform prior.


However, notable differences arise for the shape parameter $\gamma$, as seen in Fig.~\ref{fig:weibull}(c). For $\gamma < 5$, the uniform-based estimator exhibits significant variance and bias. In contrast, the Jeffreys-based estimator has considerably higher accuracy in this region, since Jeffreys prior focuses sampling efforts where the FIM is greatest. Remarkably, despite allocating fewer training samples to higher $\gamma$ regions, the Jeffreys-based estimator maintains effective generalization in those regions, showing the advantage of information-driven weighting in the parameter space.


Overall, these experiments demonstrate that sampling according to Jeffreys prior within the TS framework substantially enhances estimation accuracy and robustness for shape-dominated parameters, while maintaining comparable performance to the uniform-based estimator along directions of weaker sensitivity. This
%
also suggests that our approach may have broader applicability in synthetic data generation and parameter estimation for complex models.


\section{Conclusions} \label{sec: conclusion}
In this paper, we have proposed a MALA-based scheme designed to generate samples from Jeffreys prior.  Our approach imposes Jeffreys prior as the stationary distribution of a Langevin-based MCMC, where each update step of the Markov chain is determined by the gradient of the logarithm of the determinant of the FIM. Furthermore, for the case when this matrix cannot be computed analytically, we have employed a PF algorithm to approximate the score function, and thus the gradient of the log determinant of the FIM. We have validated our sampling scheme through several numerical examples, including one that has demonstrated how the diversity inherent in Jeffreys prior can enhance the performance of a data-driven estimator. In future work, we plan to further explore the use of this scheme for parameter estimation in dynamical systems, particularly using data-driven estimators that require minimal training data, by leveraging the sample diversity promoted by Jeffreys prior.  

\bibliographystyle{ieeetr}
\bibliography{reference}

\end{document}